\documentstyle[psfig]{espart}

\begin{document}

\begin{frontmatter}

   \title{The window Josephson junction: a coupled linear nonlinear system}
   
   \author[dresden]{A. Benabdallah},
   \author[rouen]{J.\ G. Caputo} and
   \author[crete]{N. Flytzanis}
    
   \address[dresden]{Max-Planck-Institut f\"ur Physik komplexer Systeme, 
   N\"othnitzer Stra\ss e 38, D-01187 Dresden, Germany}
   \address[rouen]{Laboratoire de Math\'ematique, INSA de Rouen,
   B.P. 8, 76131 Mont-Saint-Aignan cedex, France}
   \address[crete]{Department of Physics,  University of Crete,
   GR-71409, Heraklion, Greece.}
   \date{\today}

   \begin{abstract}

   We investigate the interface coupling between the 2D sine-Gordon equation and
   the 2D wave equation in the context of a Josephson window junction using
   a finite volume numerical method and soliton perturbation theory. The
   geometry of the domain as well as the electrical coupling parameters are
   considered. When the linear region is located at each end of the nonlinear
   domain, we derive an effective 1D model, and using soliton perturbation
   theory, compute the fixed points that can trap either a kink or antikink
   at an interface between two sine-Gordon media. This approximate 
   analysis is validated by comparing with the solution of the partial
   differential equation and describes kink motion in the 1D window junction.
   Using this, we analyze steady state kink motion 
   and derive values for the average speed in the 1D and 2D systems. 
   Finally we show how geometry and the coupling parameters can destabilize
   kink motion.
   \end{abstract}


   \end{frontmatter}

   \section{ Introduction}

   Nonlinear wave equations in one dimension that are close to integrable
   have been studied extensively for the past twenty years. The
   shift from exact integrability can be due to the presence of extra terms 
   in the equation or to an additional evolution equation. In the former case
   perturbation methods give good results. This kind of approach is far 
   more difficult
   when two such equations are coupled. In that case one needs to resort
   to extensive numerical studies in order to gain insight and develop
   approximate models. This is also true when one uses realistic boundary
   conditions, such as an impedance condition.

   \noindent Here we consider a 2D sine-Gordon equation coupled to a 2D wave
   equation via interface conditions. Such a model was originally derived
   for a so-called window Josephson junction \cite{cfd94,cfv95} between two
   superconductors where the thickness of the oxide layer is small in
   the junction area and larger in the surrounding region. The electrodynamics
   of such a system is described by the sine-Gordon equation in the window
   area \cite{Barone} where tunneling of electron pairs is possible 
   and by the wave
   equation in the surrounding passive region. This model is much more
   general and could describe the motion of dislocations in 
   a stressed inhomogeneous thin plate \cite{fk39}, in this case
   the on site potential would be non uniform. Another example 
   is the motion of 
   domain walls in a one dimensional inhomogeneous ferromagnet 
   \cite{degm84}. One could also generalize the 2D rotator array of
   \cite{rtp94}.
   From a fundamental point of view, the window junction allows to 
   study in detail the interactions between a linear and a 
   nonlinear system. The relation between the two subsystems can 
   be changed by using a different
   geometry or topology or by modifying the coupling parameters. In
   the case of the Josephson junction, the former are the junction inductance
   and capacitance per unit area.

   Let us now recall what has been done to solve this problem.
   For static solutions the longitudinal extension of the passive region
   has been shown not to play a big role, only the lateral extension 
   changes the solution \cite{cfv96}.
   For small extensions w',  the length of the problem is rescaled
   from $\lambda_J $ into 
   $\lambda_{\rm eff}  =  \sqrt{1 + 2 {w' \over w}  {L_J \over L_I}}$
   \cite{cfv96} 
   where $L_J$ (resp. $L_I$) is the inductance in the junction (resp. passive
   region). 
   This rescaling explains very well the static solutions
   in the presence of a magnetic field \cite{cfklv99}, it is a local effect
   which can also be seen for annular geometries \cite{ustinov}.
   For very large passive regions, the kink width is proportional to
   the inverse of the width of the junction, it can be very big and cause 
   the destruction of the kink due to the longitudinal junction
   boundaries\cite{cfd94}. 
   This could explain the absence of fluxon motion in window junctions 
   with large passive regions.

   \noindent Lee and co-workers \cite{Lee,lb92} investigated an infinite linear 
   superconducting strip line where no Josephson current is present. 
   They obtained the dispersion relation 
   indicating that waves can propagate in the x direction and have
   a transverse structure in y. This study done in the linear case
   does not give any indications on the effect of the linear passive
   region on kinks which are strongly nonlinear.
   Another issue is that this analysis is based on eigenmodes and 
   cannot be extended to the case when there is an external current 
   applied to the device.

   \noindent In \cite{Monaco} experiments have been conducted on Fiske 
   steps in window junctions with two different geometries a 
   lateral passive region and longitudinal passive region. 
   Lumped elements were assumed and simple models 
   were derived from which the velocity was obtained. A passive region
   placed at each end of the junction acts as a lumped capacitance and
   generates radiation. On the contrary a passive region placed along 
   the junction acts as a transmission line in parallel to the junction.
   It gives a rescaling of the Josephson penetration length and a larger
   fluxon rest mass in agreement with \cite{cfd94}. 
   Experimental results for different extensions of the passive region
   together with preliminary numerical results have been reported
   in \cite{tuk95}. These showed that fluxon motion ceases to be stable
   when the ratio of the widths of the passive region and the junction
   becomes larger than 3. The influence of the electric parameters was
   not studied.

   \noindent In a recent work one of the authors studied the case of a window
   junction with a homogeneous lateral passive region with periodic
   boundary conditions in the longitudinal direction. The motion
   of kinks in such a device occurs for velocities which can be
   calculated from the parameters of the device. In this study
   we will show that the presence of the passive region along
   the direction of propagation will impose the velocity of the
   kink.

   We have studied the problem in a systematic way by comparing
   the solution of an effective 1D problem with the 2D solution. 
   We varied both the geometrical parameters and the electric
   parameters. In a first step we considered the propagation of a kink
   across an interface where the coefficient of the kink term and the
   electric properties vary abruptly. The operator describing this
   situation is discontinuous and one needs to use a finite volume
   approximation in order to have an accurate numerical solution.
   This method based on integrating the operator on reference
   volumes enables to satisfy exactly the jump conditions at the
   interface. Using this well suited numerical method together
   with soliton perturbation techniques, we have derived simple
   models explaining the dynamics of kinks in a window junction.

   The paper is organized as follows, in section 2 we introduce
   the 2D partial differential equation model. In section 3 we
   obtain an effective 1D model for the case where the passive
   region exists only at each end of the junction. In section 4
   we study kink motion across an interface, we derive the perturbation
   equations, analyze their fixed points and validate this approach
   by comparing the solution to the one given by a numerical integration
   of the partial differential equation. Section 5 discusses kink motion
   in a 1D and 2D window junction. In section 6 we present kink motion
   instabilities in the system, consider limiting cases and give our
   concluding remarks.

   \section{ The model}

   The electrodynamics of a window junction can be described by
   two 2D arrays of inductances coupled together through an RSJ
   element containing a capacitor, the nonlinear Josephson element
   and a resistor\cite{Devoret,Likharev}. This approach is
   equivalent to a discretisation of Maxwell's equations 
   together with Josephson's constitutive equation. We then assume perfect 
   symmetry and go to the continuum
   limit to obtain the following partial differential equation for the
   evolution of the phase difference $\phi$ in a domain $\Omega$
   \begin{equation} \label{wjj}
   C \phi_{tt} -{\rm div} \left({1\over L} \nabla \phi\right) + 
   \epsilon(x,y) ( \sin(\phi) + \alpha \phi_t) = 0,
   \end{equation}
   where $C$ and $L$ are respectively the
   normalized capacitance per unit surface and inductance and 
   $\epsilon(x,y)$ is the indicator function of the junction 
   domain $\Omega_J$ (ie $\epsilon=1$ in $\Omega_J$ and 0 outside).
   The unit of space is the Josephson penetration depth and the
   unit of time the plasma frequency in the junction \cite{Barone}.

   \noindent Defining $C_I$ and $L_I$ to be respectively the normalized 
   capacitance per unit surface 
   and inductance in the passive region, Equation (\ref{wjj}) can be rewritten 
   as the system 
   \begin{equation}
   \label{wjj1}
   \frac {\partial^2 \phi}{\partial t^2} - \Delta\,\phi
   + \sin \phi + \alpha
   \frac {\partial \phi}{\partial t}= 0 \quad {\rm in} \quad \Omega_J \; ,
   \end{equation}

   \begin{equation}
   \label{wjj2}
   C_I \frac {\partial^2 \psi}{\partial t^2}
   - {1 \over L_I} \Delta\,\psi=0 \quad {\rm in} \quad \Omega 
   \backslash \Omega_J \; ,
   \end{equation}

   \noindent together with the interface conditions 
   for the phase and its normal gradient
   the surface current on the junction boundary $\partial \Omega_J$
   \begin{equation}  \label{ic}
   \psi = \phi ~~~ {\rm and} ~~~
   {1 \over L_I} {\partial \psi \over \partial n} = {\partial \phi \over 
   \partial n} 
   \end{equation}
   where $n$ is for example the exterior normal. The boundary 
   condition on the boundary of the passive region 
   represents the input of an external current or magnetic field
   \begin{equation}  \label{bc}
   {1 \over L_I} {\partial \psi \over \partial n} = f .
   \end{equation}
   \noindent Note that the jump condition (2nd relation in (\ref{ic}))
   can be obtained by integrating (\ref{wjj}) on a small surface
   overlapping the junction domain $\Omega_J$.

   In the rest of the paper we have assumed a rectangular window 
   of length $l=10$ and width $w=1$ embedded in a 
   rectangular passive region of extension $w'$ as shown in 
   Figure 1. We will not consider the 
   influence of an external magnetic field and will assume 
   the external current feed to be of
   overlap type so that the boundary conditions (\ref{bc}) become

   \begin{eqnarray}  \label{rbc}
   {1 \over L_I} {\partial \psi \over \partial y} &=& \mp {I \over 2 (l + 2 w')}
   ~~{\rm for}~~~ y = \pm ({w \over 2} + w'). \\ \nonumber
   {\partial \psi \over \partial x} &= 0 & ~~{\rm for}~~~ x = \pm ({l \over 2} + 
w').
   \end{eqnarray}

   We assumed throughout the study a small damping $\alpha=0.01$ which is
   typical of underdamped Josephson junctions.

   \section{   The 1D effective model }

   In this section we introduce a simplified model where the
   passive region exists only on the longitudinal sides of the
   junction as in the bottom right panel of Fig. 1. Then
   the functions $L$ and $\epsilon$ depend only on the variable
   x. To simplify the problem we assume a uniform boundary condition
   ${\partial \psi \over \partial y} = \mp {I \over 2 (l + 2 w')} $
   for $y = \pm ({w \over 2} + w')$. 

   We will show that this system is well described
   in the limit of a small current $I$ by a one dimensional sine-Gordon
   equation. 

   \noindent For that we write the solution of (\ref{wjj}) as 
   $\phi = \phi_I + \phi_R$ where 
   \begin{equation}  \label{phii}
   \phi_I = -{I y^2 \over 2 w (l +2 w')}
   \end{equation}
   \noindent satisfies the y boundary conditions. These also imply that
   the residual $\phi_R$ is even in y and can be expanded in a
   cosine Fourier series
   \begin{equation}  \label{phir}
   \phi_R = \sum_{n=0}^\infty A_n(x,t) \cos {2 n \pi y \over w} ~~.
   \end{equation}
   Inserting (\ref{phii}) and (\ref{phir}) into (\ref{wjj}) we obtain
   $$ C[A_0{_{tt}} + A_{1_{tt}} \cos {2 \pi y \over w}+ ... ]
   -\left[ {1 \over L}(A_{0_x} + A_{1_x} \cos {2 \pi y \over w}+ ...)\right]_x
   -\left[ {1 \over L}\left(-{I y \over w ( l + 2 w')}\right. \right.  $$
   $$ \left. \left.
   - {2 \pi \over w} A_1 \sin {2 \pi y \over w}+ ...  \right)  \right]_y 
   + \epsilon(x) [\sin(\phi_I + A_0 + A_1 \cos {2 \pi y \over w}+ ... )  
   + \alpha ( A_{0_t} + A_{1_t} \cos {2 \pi y \over w}+ ...) ] = 0 ~~. $$
     
   \noindent We get the evolution of $A_0$ by integrating the equation 
   above in $y$ and dividing by $w$,
   $$
   C A_{0_{tt}} - \left({1 \over L} A_{0_x}\right)_x + {I \over L w ( l + 2 w')} 
   +  \epsilon(x) \left[ {1 \over w} \int_{- w \over 2}^{w \over 2} 
   \sin(\phi_I + A_0 + A_1 \cos {2 \pi y \over w}+ ... ) dy +\alpha 
A_{0_t}\right] =
   0~~.$$
   \noindent We write 
   $$\sin(\phi_I + A_0 + A_1 \cos {2 \pi y \over w}+ ... ) \approx 
   \sin(\phi_I + A_0) + A_1 \cos {2 \pi y \over w} \cos(\phi_I + A_0)$$
   because $|A_1| << |A_0|$. First we consider the integral of 
   the first term $\sin(\phi_I + A_0)$
   $$\int_{- w \over 2}^{w \over 2} \sin(\phi_I + A_0) dy =
   \cos(A_0) \int_{- w \over 2}^{w \over 2} dy \sin(\phi_I) 
   + \sin(A_0) \int_{- w \over 2}^{w \over 2} dy \cos(\phi_I)$$
   \noindent The first integral on the right hand side can be written
   $$\int_{- w \over 2}^{w \over 2} dy \sin(\phi_I) = w
   \int_0^1 d\zeta \sin( -{I w \over 8(l+ 2 w')} \zeta^2)$$
   \noindent so that if $-{I w \over 8(l+ 2 w')} << 1$ the sine
   can be linearized and the cosine taken equal to 1 leading to
   $$\int_{- w \over 2}^{w \over 2} \sin(\phi_I + A_0) dy = 
   -{I w \over 24 (l+ 2 w')} \cos(A_0) + w \sin(A_0)   ~~.$$

   \noindent The terms in $A_1$ are
   $$ A_1 \cos(A_0) \int_{- w \over 2}^{w \over 2} dy \cos(\phi_I )
   \cos {2 \pi y \over w} 
   -A_1 \sin(A_0) \int_{- w \over 2}^{w \over 2} dy \sin(\phi_I )
   \cos {2 \pi y \over w}  ~~.$$
   Both terms can be treated following the same approximations
   as above so that
   $$\int_{- w \over 2}^{w \over 2} dy \cos(\phi_I )
   \cos {2 \pi y \over w} \approx {w \over \pi}  ~~~;~~~
   \int_{- w \over 2}^{w \over 2} dy \sin(\phi_I )
   \cos {2 \pi y \over w} = -{I w^2 \over 4 \pi^3 (l+ 2 w') } ~~.$$

   \noindent Combining all contributions, we obtain for the evolution of
   $A_0$
   \begin{equation} \label{a0}
   C A_{0_{tt}} - \left({1 \over L} A_{0_x}\right)_x + {I \over L w ( l + 2 w')}
   +  \epsilon(x) [ \sin( A_0) - 
   \end{equation}
   $${I w \over 24 (l+ 2 w') } \cos(A_0)
   +A_1 ({w \over \pi} \cos(A_0) - {I w^2 \over 4 \pi^3 (l+ 2 w') } \sin(A_0) )
   + \alpha A_{0_t}] =0 .
   $$

   \noindent For the values of the geometric parameters that we have taken
   $l=10,w=1, |w'| \le 10$ and if we assume $I <1$ which is the case for
   a zero field step then the terms $ {I w \over 24 (l+ 2 w') } , {w \over \pi},
   {I w^2 \over 4 \pi^3 (l+ 2 w') }$ can be neglected so that we are left with
   the following sine-Gordon equation for $A_0$ (dropping the $0$)
   \begin{equation} \label{a0f}
   C A_{tt} - \left({1 \over L } A_x\right)_x + \gamma(x)
   +  \epsilon(x) [ \sin( A) +  \alpha A_t] =0 ,
   \end{equation}
   \noindent where $\gamma(x) = {I \over L(x) w ( l + 2 w')}$.
   This approach is validated for a homogeneous 2D sine-Gordon
   equation by the numerical evidence
   provided by Eilbeck and al \cite{eilbeck} that when 
   $(I/{8 {\cal L}} \ll  1)$, the phase is almost uniform in the
   $y$ direction.

    \section{Motion of a kink across a 1D interface}
   
   Before considering the motion of a  kink in a window junction, we will
   study the simpler problem of an interface between two
   media with different values of the parameters $\{L(x),\;C(x),\;
   \epsilon\}$. The propagation of a soliton across an interface
   has been studied by many authors \cite{sso87}, \cite{km89}, \cite{kc91}. A 
   pioneering work was done by
   Aceves et al \cite{Aceves} for the motion of a nonlinear Schroedinger soliton
   between two media of different Kerr indices. It was shown
   that the soliton perturbation equations give a qualitatively correct
   description even when the nonlinearity of the second medium 
   is very small. We will follow the same route and 
   derive adiabatic equations for the soliton 
   parameters. These are in principle only valid for small perturbations
   but some ideas can be obtained by taking the limit $\epsilon \rightarrow 0$.
   In this way one can understand qualitatively the motion of a kink from 
   a junction to a passive region.

   Following \cite{sso87}, we introduce the change of variable
   \begin{equation} \label{dzdx}
   {\rm d}z = \sqrt{L(x)} {\rm d}x\;,
   \end{equation}
   \noindent which makes the perturbation theory regular in the
   sense that the solution has the same form on the left and right
   of the interface. In principle the terms $C A_{tt}$ and $\epsilon \sin A$
   make the perturbation theory singular, however we are mostly interested
   in the fixed points that exist due to the interfaces, The first term
   will not influence these, nor their stability. We expect this
   description to hold if the wave remains close to the interface.

   Then we write equation (\ref{a0f}) in standard form 
   \begin{eqnarray}
   \nonumber
   A_{tt} - A_{zz}+\sin (A) = &-&\gamma (x) 
   -\alpha A_t 
   +(1-\epsilon(x)) \sin (A )
   +(1-C(x))A_{tt} \\ \label{sgt}
   &-& \frac 12\partial_z\left[{\rm ln }L(x)\right]
   A_z \; .
   \end{eqnarray}
   To simplify the calculations we assumed a uniform damping $\alpha$.
   To model the experimental situation of \cite{tuk95} for example,
   it is natural to assume the following dependencies for the 
   parameters $L,\, C\, {\rm and}\, \epsilon$ 
   \[
   L(x)=\left\{
   \begin{array}{c}
   1\;\;{\rm if }\;\;x < 0 \\
   L_I\;\;{\rm if }\;\;x > 0
   \end{array}
   \right.
   \;\;\;, \;\;\;
   C(x)=\left\{
   \begin{array}{c}
   1\;\;{\rm if }\;\;x < 0 \\
   C_I\;\;{\rm if }\;\;x > 0
   \end{array}
   \right.
   \;\;\;{\rm and} \;\;\;
   \epsilon(x)=\left\{
   \begin{array}{c}
   1\;\;{\rm if }\;\;x < 0 \\
   \epsilon_0\;\;{\rm if }\;\;x > 0 \; .
   \end{array}
   \right.
   \]
   \noindent With this type of distribution of the 
   inductance $L$ and using
   (\ref{dzdx}) we obtain the differential relation 
   $ \partial_z \left[{\rm ln} L(x)\right]=({\rm ln}(L_I))\delta(z)\; ,$
   where $\delta$ is the Dirac delta function.\\
   \noindent Now we rearrange the right hand side of Eq. (\ref{sgt}) as the perturbation 
   term 
   \begin{equation}
   \label{eff}
   \epsilon f= -\gamma_J -\alpha A_t
   +\left[ -(\gamma_I -\gamma_J)
   -\mu_0\sin A + (1 -C_I)A_{tt}
   \right]H(z)
   -\frac {{\rm ln}(L_I)}2\delta(z)A_z \; ,
   \end{equation}
   where $\mu_0 = \epsilon_0 -1$, $\gamma_I =
   {I \over L_I w ( l + 2 w')}$, 
   $\gamma_J = {I \over w ( l + 2 w')}$ and $H$ is usual Heaviside function.

   Note that the functions $C$
   and $\epsilon$ are piece-wise constants and therefore 
   there is no different re-scaling on each side of the
   interface as we go from the variable $x$ to $z$.
   Having set up the problem (\ref{sgt}), we assume 
   as usual that the
   soliton parameters are slowly modulated and use perturbation
   theory \cite{ms78} to derive their evolution.

   \noindent We choose the kink ansatz
   \[
   A_a(z,t)=4\tan ^{-1}\exp \sigma \left(\frac{z-Z(t)}{\sqrt{1-v^2}}
   \right )\, ,
   \]
   where $\displaystyle{Z(t)=\int_0^tv(t^{\prime })
   {\rm  d}t^{\prime } + z_0(t)}$ and $\sigma = 1$ (resp. $-1$) for
   a kink (resp. antikink).\\
   Then $(Z(t),v(t))$ are solutions of the differential system 
   \begin{eqnarray}
   \nonumber
   \frac{dZ}{dt}&=&v
   -\frac{\mu_0 v(1-v^2)}2\left( 1+\tanh
   \left(\frac Z{\sqrt{1-v^2}}\right)\right)    
   - \frac {Zv}{\sqrt{1-v^2}}
   \left( \frac{\mu_0 (1-v^2)}2 + \frac {{\rm ln}(L_I)}4\right)\\ \nonumber
   &\times & 
    {\rm sech}^2\left(\frac Z{\sqrt{1-v^2}}\right) +
   \frac{(1-C_I)v^2\sqrt{1-v^2}}4\left( 1-\frac Z{\sqrt{1-v^2}}\right)
    \left[ 1+ \tanh \left(\frac Z{\sqrt{1-v^2}}\right) \right]\\ \label{odez} 
   &-&{v \over 4} (\gamma_I - \gamma_J) \sqrt{1-v^2} 2 G\; ,
   \end{eqnarray}
   \begin{eqnarray}
   \nonumber
   \frac{dv}{dt} &=&
   \frac{\pi \sigma (\gamma_I+\gamma_J) \left( 1-v^2\right) ^{\frac 32}}8
   -\alpha v(1-v^2)
   -\frac {\sqrt{1-v^2}}4
   \left(\mu_0 (1-v^2) - {\rm ln}{(L_I)}\right) \\ \label{odev} 
   & \times & {\rm sech}^2 \left(\frac Z{\sqrt{1-v^2}}\right) + 
   \frac{(1-C_I)v^2\sqrt{1-v^2}}4
   {\rm sech}^2 \left(\frac Z{\sqrt{1-v^2}}\right)
     \,  ,
   \end{eqnarray}

   \noindent where $ 2 G = 1.83 193 11 88..$ \cite{Gradstein}. If
   we assume $\epsilon\equiv 1,$ $\alpha=I=0$ and $C_I=1$ these equations 
   are exactly the ones derived in \cite{sso87}.
   
   In the absence of the current and the damping, the equations 
   (\ref{odez}-\ref{odev}) derive from the Hamiltonian,
   \begin{equation}
   \label{hamm}
   H=\int_{-\infty}^{+\infty} L(z)^{-1/2} \left[
   \frac{C(z)}2\left(\frac{\partial \varphi} {\partial t}\right)^2
   +\frac12\left(\frac{\partial \varphi} {\partial z}\right)^2
   +\epsilon(z)(1-\cos \varphi)\right] \;{\rm d}z\; .
   \end{equation}
   \noindent If we assume the fluxon to have velocity $v_l$ for 
   $z \ll 0 $, its energy is $H_l(v_l) = \frac 8{(1 - v_l^2)^{1/2}}$.
   If it crosses into the passive medium $z \gg 0$, it will
   have a velocity $v_r$ and energy $H_r(v_r)$ 
   \begin{equation}
   H_r(v_r) = \frac {4 v_r^2(C_I - \epsilon_0) + 4(1 + \epsilon_0)}
   {L_I^{1/2}(1 - v_r^2)^{1/2}} ,
   \end{equation}
   and one can obtain the condition on the velocity $v_l$ so 
   that the left coming soliton crosses into the right hand medium as 
   $H_l(v_l) \ge H_r(v_r=0)$ which implies  
   $v_l^2 > 1 - \frac {4L_I}{(1 + \epsilon_0)^2} \; $.
   One can then compute $v_r$ by identifying $H_l$ and $H_r$.
   If $v_l <\sqrt{1 - \frac {4L_I}{(1 + \epsilon_0)^2}}$ the fluxon is
   reflected with velocity $-v_l$.

  \subsection{Existence and stability of fixed points}

   From the above equations, we obtain two fixed points symmetrically 
   placed with respect to the interface $z=0$
   \begin{equation}
   \label{ptf}
   Z_{\pm} = {\rm atanh}~~ \eta_\pm \equiv
   {\rm atanh}~~ \pm \left(1+\frac{\pi \gamma \sigma}
   {\epsilon_0 -1 - \log(L_I)} \right)^{\frac 12}   ,
   \end{equation}
   \noindent where $\gamma = { (\gamma_I + \gamma_J) \over 2}$.

   The expression indicates that at least two ingredients
   are necessary for soliton trapping at a finite distance
   from the interface, a DC current $\gamma$, a jump
   in the coefficient of the sine nonlinearity $\epsilon$
   (the critical current density in the Josephson model)
   and an inductance jump $L_I$. Trapping has been predicted
   with the first two features in the works \cite{km89,kc91}.
   Notice also that the existence and position of the fixed points 
   is completely independent of the damping $\alpha$ and capacity $C_I$ in
   the passive region.

   Let us consider the conditions for existence of 
   these fixed points. For that we separate the cases of a 
   kink or an antikink. For a kink $\sigma=1$, the fixed points 
   exist if $ \mu_0 -log(L_I) -\pi\gamma \ge 0$ or 
   $L_I \le e^{\mu_0 -\pi \gamma}$
   while for the antikink $\sigma = -1$ they exist if 
   $L_I \ge e^{\mu_0 + \pi \gamma}$.

   To investigate the stability of these fixed points we compute the
   Jacobian of the evolution equations (\ref{odez}-\ref{odev}) and estimate its
   eigenvalues $\lambda$ at each of the fixed points. The characteristic 
   equation is
   $$\lambda^2 + \lambda \alpha -{\pi \gamma \sigma \eta_\pm \over 2}
   [1 -{\mu_0 \over 2}(1+\eta_\pm) - {\pi \over 2} Z_\pm \gamma \sigma 
   { \mu_0 +{\log(L_I) \over 2} \over \mu_0-\log L_I} ] = 0 .$$

   \noindent Let us fix the type of wave to be an antikink. Then we 
   find that the discriminant for 
   the fixed point $Z_+$ is negative for any value of the current and
   tends to zero as the current tends to zero. This fixed point is then
   stable because the eigenvalues are complex conjugate and their real
   part $-\alpha /2 $ is negative. On the other hand the fixed point
   $Z_-$ is unstable because the discriminant is always positive
   and the eigenvalues are of opposite signs. 

   \noindent If we had considered as for the 2D problem that dissipation is
   absent on the right hand medium, we would have obtained half
   the damping term in the equation for $d v / dt$  and an additional
   term $\alpha {\epsilon \over 2} \sqrt{1-v^2} v^2 \sigma \log(2)$
   in the equation for $d Z / dt$. The first term would affect the
   stability, it is equivalent to halving the overall damping term.

   \noindent The existence and stability for both
   kink and antikink waves is summarized in Table 1.

   \begin{figure}

   \caption{Table 1: existence and stability of the fixed points
    $Z_\pm$} 

   \centerline{
    \begin{tabular}{|l|c|c|r|}
    \hline
    {\it } & $L_I \le e^{\epsilon_0 -1 -\pi\gamma}$  & 
    $e^{\epsilon_0 -1 -\pi\gamma} \le L_I \le e^{\epsilon_0 -1 + \pi\gamma}$&  
    $e^{\epsilon_0 -1 + \pi\gamma} \le L_I$  \\ \hline
                              & & & 2 fixed points $Z_\pm$    \\  
    antikink ($\sigma = -1$) & no fixed points & no fixed points & $Z_+$ is 
stable  \\  
                           & &  & $Z_-$ is unstable      \\  \hline
                        &2 fixed points $Z_\pm$ &  &    \\ 
    kink ($\sigma = 1$) & $Z_+$ is unstable & no fixed points &no fixed points   
 \\ 
                        & $Z_-$ is stable&  &    \\ \hline
    \end{tabular}
    }
   \label{table1}
   \end{figure}

   \noindent This study that such an interface enables to trap a certain
   type of wave. For large $L_I$, antikinks are trapped at a 
   fixed point $X_+= Z_+/\sqrt(L_I)$ in the passive region 
   while kinks remain free. On
   the contrary when $L_I$ is small, kinks are trapped at the fixed point
   $X_-=Z_-$. If the interface is now such that the passive region is for $z<0$
   and the junction for $z>0$, then the second column and the last column of 
   the table should be permutated and the stability properties exchanged. 
   This provides a qualitative understanding of the window device which 
   contains two such interfaces.

   \subsection{Validation of the perturbation theory}

   In order to test these predictions we have compared the solution
   of the partial differential equation (\ref{a0f}) with the solution
   of the perturbation equations (\ref{odez}-\ref{odev}). Note that the former
   is a generalized operator with discontinuities in the coefficients
   so that the solution is continuous but exhibits jumps of its derivative 
   at the junction interface, To obtain it we introduce the finite
   volume method used for the treatment of hyperbolic equations \cite{volfin}.
   The idea is to integrate the partial differential equation over intervals
   around each discretisation point where the solution is assumed constant.
   In this way the jump conditions are
   exactly respected. The details of the implementation for both the 1D
   and 2D cases are given respectively in appendices A and B.

   Fig. 2 shows the phase plane in the original position-velocity coordinates
   $ (X,u)$ for an antikink wave for two different
   interfaces with $\epsilon_0=0.6, \gamma = 0.02$ and $L_I = 1$ (top panel)
   and $L_I = 1.5$ (bottom panel). The solution of the perturbation equations
   is given in full line while the solution of the partial differential equation
   is given by the crosses. A least square procedure has been used to estimate
   the position and velocity of the wave. For both cases we obtain a
   good overall agreement between the solution of the partial differential
   equation and the perturbation approach, showing that the fixed point traps
   the antikink even for a large initial velocity ($u=0.5$). For the bottom
   panel where $L_I = 1.5$ notice the jump in the phase gradient and the fact
   that the fixed point is close to the interface as expected.

   We now consider that the nonlinear term on the right hand side is very small.
   Fig. 3 shows $X(t)$ and $u(t)$ for an antikink in the case $L_I=1$ and
   $\epsilon_0=0.1$. For such a strong perturbation, we obtain a qualitative
   agreement for both the position and velocity of the wave between the
   solutions of (\ref{a0f}) and (\ref{odez}-\ref{odev}). Both the numerical solution and
   the perturbation method show the existence of a stable fixed point located in
   the region $\epsilon(x) = \epsilon_0 < 1$. We therefore expect (\ref{odez}-\ref{odev})
   to provide qualitative results even for $\epsilon_0 \rightarrow 0$ i.e. the
   nonlinear-linear interface, It is then possible to use these arguments to 
describe
   the motion of a kink in a window junction where the passive region exists
   on both sides of the Josephson junction.

   \section{Motion of a kink in a window junction} 

   \subsection{The perturbation theory}

   As shown in the perturbation equations (\ref{odez}-\ref{odev}) the motion of
   a kink is due to the injection of direct current in the device, described by 
the
   $\gamma$ term. In a pure Josephson junction, this gives rise to the
   well-known Zero (magnetic) Field Steps in the current voltage IV 
characteristics.
   The presence of passive regions where no sine term is present and for which 
there
   is an inductance jump will affect the kink motion. We will concentrate in 
this
   study on the features of this motion. The study of the IV characteristics and
   its associated zero field step will be presented in \cite{bc00j}.

   The study conducted above for the case of a single interface provides a way
   to give a simple description of the fluxon motion in a one dimensional
   window junction (bottom left panel in Fig. 1). In that case the current
   density is $\gamma_{1dw} = I(1 + 1/L_I) /(2(l+2 w'))$ and the position
   of the fixed points is
   \begin{equation}
   \label{ptfx}
   X_{\pm} = \frac{1}{L_I} {\rm atanh}~~ \pm \left(1+\frac{\pi \gamma \sigma}
   {\epsilon_0 -1 - \log(L_I)} \right)^{\frac 12}   ,
   \end{equation}
   This quantity depends weakly on both the current $I$ and the inductance
   in the passive region $L_I$. For the parameters used here $\epsilon_0 = 0$,
   $l=10$, the fixed point is inside the domain as soon as $w' \ge 2$.

   Fig. 4 shows the schematic motion of a wave in such a device, in the limiting
   cases $L_I >> 1$ (top panel) and $L_I << 1$ (bottom panel). The antikink wave
   propagates towards the right and gets reflected as a kink at the right 
boundary
   of the device. When $L_I >> 1$, the antikink finds two fixed points on each 
side
   of the interface, identified by large or small circles depending whether the
   fixed point is stable or unstable. The antikink is slowed down and can be 
trapped
   at the fixed point located to the left of the left interface. The kink can 
become
   trapped at the fixed point of the right interface. The situation is reversed 
for
   $L_I << 1$. We see that these points will affect the motion and could hinder 
wave
   motion.

   \subsection{Numerical study}

   To confirm the role of the fixed points at the interfaces we carried
   out a systematic study of the kink motion in both a 1D and a 2D 
   configuration. The numerical procedure in both cases is to discretize
   the spatial operator using the finite volume approximation which allows
   to preserve the interface conditions (see Appendix A and B). 
   The temporal part is then advanced
   using the Dormand and Prince ordinary differential equation solver DOPRI5
   implemented by Hairer and Norsett \cite{hn87}.
   The initial condition is a static kink for the 1D junction and the
   static solution \cite{cfv96} in the case of a 2D window junction.
   The wave is then accelerated by the injection of direct current
   through the boundary conditions (\ref{rbc}). It will reach its limiting
   velocity which can be obtained by simple arguments.
   
   \noindent Figure 5 presents the motion of a kink at its limiting velocity
   in a 1D window junction. The top and middle
   panels show respectively the position $X(t)$ and velocity $u(t)$
   vs. time while the bottom panel shows the phase plane $(X,u)$.
   The electric parameters are $L_I=C_I=2$ so that the velocity
   of the linear waves in the passive region is $v_I= 1/\sqrt{L_I C_I}=0.5$.
   One sees that the kink velocity is close to 1 in the junction
   region while it is about 0.5 in the passive region, so the kink 
   adapts its speed to the region it travels in. At this point it is
   interesting to remark that the kink continues to exist even in
   the regions where there is no sine term. The whole motion and
   the voltage observed are then given by the expression
   \begin{equation}
   \label{v1d}
   V_{1D} \equiv <\phi_t>= { \Delta \phi \over \Delta t}= 
      { 2 \pi \over 2 w' / v_I + l} .
   \end{equation}

   The motion of a kink in a 2D window junction is different as
   can be seen from Figure 6 which shows a contour plot of the 
   phase $\varphi(x,y,t)$ for 4 successive values of time. The
   parameters are the same as in Figure 5. In this case no
   significant change is observed in the speed of the wave in
   the junction and passive regions. The voltage observed 
   indicates that the velocity is everywhere equal to the
   one in the passive region $v_I= 1/\sqrt{L_I C_I}$. There
   is a simple explanation for this, the soliton velocity 
   is a free parameter for the sine Gordon equation. In the
   linear region waves can only travel at velocity $v_I$.
   The dressed kink which carries a significant part of its
   energy in the passive region adapts its speed to $v_I$.
   The voltage in this case is
   \begin{equation}
   \label{v2d}
   V_{2D} = { 2 \pi v_I \over 2 w' + l} .
   \end{equation}
   \noindent For a large extension $w'$ of the passive region or
   values of the electric parameters non equal to 1, the kink motion can 
   break down. In the next section, we give a few examples of this
   phenomenon.

   \section{Instabilities and limiting cases}

   In our procedure, we take as initial condition a static
   kink and accelerate it by injecting current via the boundary condition
   (\ref{bc}). We present now a few situations where this procedure led
   to the break down of the kink and another type of dynamical behavior
   occurred. The situation is simpler for the 1D window junction and
   we will present this first.

   In the 1D case, the kink (antikink) can be trapped at the fixed point 
   on the right (left) end interface. This occurs because as the width of 
   the passive region $w'$ is increased, the current density is decreased
   so that there is less driving force to overcome the potential barrier
   created by the stable fixed point. For the junction geometry that we chose
   $l=10, w=1$ kink motion breakdown occurs for  $w'\approx 10 $.

   \noindent The electrical parameters act differently. To see this
   consider the equations of the problem
   \begin{equation}
   \label{wj1d1}
   \frac {\partial^2 \phi}{\partial t^2} 
   - \frac {\partial^2 \phi}{\partial x^2}    + \sin \phi +\gamma + \alpha
   \frac {\partial \phi}{\partial t}= 0 \quad |x| \le {l \over 2} \; ,
   \end{equation}
   \begin{equation}
   \label{wj1d2}
   C_I \frac {\partial^2 \psi}{\partial t^2}
   - {1 \over L_I} \frac {\partial^2 \psi}{\partial x^2} +\gamma=0 
   \quad {l \over 2} \le |x| \le {l \over 2} + w'\; ,
   \end{equation}
   \noindent together with the interface condition at $|x|={l \over 2}$
   $$\phi = \psi ~~~, {1 \over L_I} \frac {\partial \psi}{\partial x}
   = \frac {\partial \phi}{\partial x}~~,$$
   \noindent and homogeneous boundary conditions 
    $\frac {\partial \psi}{\partial x}|_{x = \pm (l/2 + w')} = 0$.

   \noindent When $C_I$ or $L_I$ are large we can consider their inverse
   as a small parameter and use perturbation theory to get some estimates.
   Let us first consider the case $C_I \gg 1$. Then from equation (\ref{wj1d2})
   we get $\frac {\partial^2 \psi}{\partial t^2}=0$ so that 
   $\frac {\partial\psi}{\partial t}=V$ is a constant. We then write
   $\psi = V t + f(x)$ so that the (\ref{wj1d2}) becomes 
   \[
 - \frac{d^2 f}{ dx^2}  +\gamma L_I =0\, , \]

   \noindent which yields $\frac{d f}{ dx}= \gamma L_I (x-l/2 -w')$ for
   $l/2 \le x \le l/2 + w' $ and 
   $\psi = \gamma L_I (x-l/2 -w')^2 /2+ V t + Cte$. 
   This fixes the boundary conditions for the first equation. When $L_I$
   is large $\frac {\partial \phi}{\partial x}|_{x=\pm l/2} \rightarrow 0$
   so that kink motion is possible. On the contrary, kink motion
   becomes impossible when $L_I <<1$ because then 
   $\frac {\partial \phi}{\partial x}|_{x=\pm l/2} \rightarrow \infty$.

   Let us now consider the two dimensional situation. For large passive
   regions the static kink solution becomes stretched and occupies the
   whole junction \cite{cfd94}. It is then difficult to accelerate and 
   generally it breaks up after a few round trips. Figure 7
   shows contour plots of the phase $\varphi(x,y,t)$, numerical 
   solution of (\ref{wjj}) for $t=0,5, 6$ and 8 clockwise starting 
   from the top left panel, for a large passive width $w'=7$. 
   One sees that the kink breaks up and leads to a radial 
   phase distribution centered on the junction.
   The wave equation then dominates the dynamics and the average phase
   increases with time. 

   Another type of instability is due to the electrical parameters $L_I$
   and $C_I$, especially the former because it acts on the interface
   condition in addition to the operator. For example consider $L_I = 10^{-4}$
   and $w'=2$, the case of Figure 8 where the phase $\varphi(x,y,t)$ is shown
   for $t=2, 10$ and 20 from left to right. The top panels show the 
   contours while the bottom ones show the three dimensional plots.
   The phase tends to obey a Laplace equation in the passive region and
   there the kink leads to a uniform phase distribution. The junction is
   strongly coupled to the passive region because the gradient at the interface
   is very large. This causes the break-up of the initial kink and 
   leads to a radial solution with small oscillations. 
   
   \noindent A very large inductance leads to the creation of a boundary layer
   along the top and bottom edges of the device as shown in Figure 9
   which presents two snapshots for successive times. In this case
   $L_I=10^4$. At the interface the gradient of $\varphi$ is close to
   zero, so that the phase is constant in the junction. Notice
   the strong gradients on the top and bottom sides of the passive region.

   For $w'=2$, we have found in the $(L_I, C_I)$ plane, the
   region of instability of the kink as shown in Figure 10.
   We plotted in the $(L_I, C_I)$ plane, the hyperbolas corresponding
   to three values of velocity in the passive region, $v_I= 0.5, 1$ and 2. 
   The signs $(*)$ $ (\times)$ and $(+)$ indicate where we found 
   a stable kink motion. We have isolated by the solid curve the stability
   region in the $(L_I\,,C_I)$ plane. As expected the instability is 
   due essentially to the inductance $L_I$ so the kink is stable in the 
   domain $0.1 \leq L_I \leq 2$.  In this interval we always observed
   kink motion independently of the value of the capacity $C_I$.
   For example for a capacity $C_I=10^{4}$ and $L_I=1$, we obtain a 
   stable kink motion. The voltage observed $V_{\rm num} 2D=
   4.510 \times 10^{-3}$ is in good agreement with the one given by 
   (\ref{v2d}), $V_{\rm 2D}= 4.5 \times 10^{-3}$. 
   Then we conclude that the region of stability of the kink in the window
   problem is
   \begin{equation}
   \label{reg_stab}
   D_{\rm stability}=\lbrace 0.1 \leq L_I \leq 2  \rbrace  \, .
   \end{equation}
   At this time we do not have an explanation for this treshold value
   $L_I=2$.

   From a general point of view this study has shown the importance
   of the perturbation approach to gain insight into a complicated
   wave problem. Another determining factor was the use of the finite
   volume discretisation to solve the inhomogeneous partial differential
   equation.
   
   {\bf Acknowledgments}

   The Authors thank A. C. Scott, A. Ustinov, and S. Flach for 
   helpful discussions. A. B. thank the department of Math\'ematiques 
   de l'INSA de Rouen. J. G. C. thank the Max Planck institute for the Physics 
   of complex systems, Dresden, for its hospitality during a short visit. 
   This work was supported by the European Union under the RTN project 
   LOCNET HPRNCT-1999-00163.

   {\bf APPENDIX A: The finite volume formulation in 1D}
   
   We introduce new functions for the 1D window equation (\ref{a0f}) 
   \[ u= A_t \quad{\rm and}\quad v= A_x  \; ,\] 
   which becomes the system
   \begin{equation}
   \label{cf}
   \frac {\partial U}{\partial t} + \frac {\partial ({\cal A} U)}
   {\partial x} = {\cal B}(U,A) \; ,
   \end{equation}
   where
   \[ U = \left [\begin{array}{cc} 
   u \\
   v \end{array} \right ] ,\, {\cal A}(x)= \left [
   \begin{array}{cc} 
   0  & -\frac 1 {L(x) C(x)}\\
   -1 &  0 \end{array} \right ],
   \] 
   and \[ 
   {\cal B}(U,A)= \left [
   \begin{array}{c} \displaystyle{
   -\frac {\epsilon(x)}{C(x)}(\alpha u + \sin A) - \frac \gamma {C(x)} -
   \frac {\partial } {\partial x} \left( \frac 1 {C(x)} \right) \frac 
    1 {L(x)}v} \\
    u
   \end{array} \right ] \; . \]
   The matrix ${\cal A} $ has two real eigenvalues 
   $ \lambda_{1,2}=\pm \sqrt{\frac 1{L(x)C(x)}}\; $ thus the system 
   (\ref{cf}) is strictly hyperbolic with a source term.
   Due to the discontinuities of the functions $C(x),L(x)$ and $\epsilon(x)$ 
   at the interfaces, the numerical integration of the Eq. (\ref{cf}) 
   using finite differences cannot be done. 
   We therefore turn to the finite volume method \cite{volfin}, whose 
   principle is first to consider a partition of 
   the passive and junction domain in cells
   \[
   V^{i,j}_{k}=\left \{ x \,| \,x_k-\frac{h_{i,j}}2 \leq x < x_k + 
\frac{h_{i,j}}2\right \}
   \, \; ,
   \]
   where $x_k$ are collocation points defined by
   \begin{eqnarray}
   \nonumber
   x_k&=&(k-\frac 12)h_i \,\, {\rm in ~~passive~~ region}\; ,\\ \label{xk}
   x_k&=&(k-1)h_j  \,\, {\rm in~~junction~~ region} \; .  \nonumber
   \end{eqnarray}
   Here $h_i$ and $h_j$ are the space mesh-length, defined by 
   \[ h_i=\frac{w'}{n_i+\frac 12} \; ,h_j=\frac{l}{n_j-1} \; ,
   \]
   where $n_i$ and $n_j$ respectively are the number of discretization
   in the passive and junction domain. The total number points of the whole 
   domain is $2n_i + n_j$.
   Notice that the points $x_k$ such $k=n_i$ and $k=n_i+n_j$
   correspond to the real interface points $x=-l/2$ and $x=l/2$.\\    
   In a second step we assume the solution of Eq. (\ref{cf})
   to be constant in each cell $V^{i,j}_{k}$. We integrate Eq. (\ref{cf})
   over each cell 
   \begin{equation}
   \label{it1}
   \int_{V^{i,j}_{k}}
   \frac{\partial U}{\partial t} + \frac {\partial ({\cal A} U)}{\partial 
x}\;{\rm d}x
   = \int_{V^{i,j}_{k}} {\cal B} \;{\rm d}x,
   \end{equation}
   and obtain
   \begin{eqnarray}
   \nonumber
   \mu(V^{i,j}_{k}) \frac{{\rm d} U_k}{{\rm d} t} +\left[ {\cal A} U 
\right]_{\partial
   V^{i,j}_{k}}=
   \int_{V^{i,j}_{k}} {\cal B}(U,A) \;{\rm d}x + {\cal O}(h^2) \; ,
   \end{eqnarray}
   where $U_k = U(x)$  for all $x \in V^{i,j}_k$.

   We therefore obtain the following system of ordinary differential
   equations
   {\bf In the passive region}
   \begin{equation}
   \frac{{\rm d}}{{\rm d} t}
   \left [ \begin{array}{cc}
   u_k\\
   v_k \end{array} \right ] = \left [
   \begin{array}{cc}
   \frac 1{L_I C_I h_i} \left(v_h({x_k +\frac{h_i}2}) -
    v_h({x_k -\frac{h_i}2})\right) \\
   -\frac {u_h\left({x_k +\frac{h_i}2}\right) 
   - u_h\left({x_k -\frac{h_i}2}\right)}{h_i} \end{array} \right ] -
   \left [ \begin{array}{cc}
   \frac \gamma {C_I} \\
    0
   \end{array} \right ] \; ,
   \end{equation}
   which in terms of $A$ yields
   \begin{eqnarray}
   \label{shi}
   \left\{
   \begin{array}{cc}
   \displaystyle{\frac {{\rm d} A_k}{{\rm d} t}= u_k}&\\
   \\
   \;\;\;\;\displaystyle{\frac {{\rm d} u_k}{{\rm d} t}= -\frac \gamma {C_I}}
   +&\displaystyle{\frac{A_{k+1}-2A_k+A_{k-1}}{L_I C_Ih_i^2} \; .
   }
   \end{array}
   \right.
   \end{eqnarray}

   {\bf In the junction domain we obtain}
   \begin{equation}
   \frac{\partial}{\partial t}
   \left [ \begin{array}{cc}
   u_k\\
   v_k \end{array} \right ] = \left [
   \begin{array}{cc}
    \frac 1{h_j} \left(v_h({x_k +\frac{h_j}2}) 
    - v_h({x_k -\frac{h_j}2})\right) \\
    -\frac {u_h\left({x_k +\frac{h_j}2}\right) 
    - u_h\left({x_k -\frac{h_j}2}\right)}{h_j}
   \end{array} \right ] - \left [
   \begin{array}{cc}
   \sin \varphi_k + \alpha u_k +  \gamma\\
   0 \end{array} \right ] \; ,
   \end{equation}
   which in terms of $A$ implies
   \begin{eqnarray}
   \label{shj}
   \left \{
   \begin{array}{cc}
   \displaystyle{\frac {{\rm d} A_k}{{\rm d} t}= u_k} & \\
   \\
   \;\;\;\displaystyle{\frac {{\rm d} u_k}{{\rm d} t}=-\alpha u_k}&
   -\;\;\;\displaystyle{\sin A_k -\gamma+ \frac{A_{k+1}-2A_k
   +A_{k-1}}{h_j^2}\; .}
   \end{array} \right.
   \end{eqnarray}
   
   {\bf At the interface we get}
   \begin{eqnarray}
   \nonumber \left \{ \frac {h_i}2 \left[
   \begin{array}{cc}
   C_I & 0 \\
   \\
   0   & 1 \end{array} \right] + \frac {h_j}2 \left[
   \begin{array}{cc}
   1 & 0 \\
   \\
   0   & 1 \end{array} \right] \right \}
   \frac{\partial }{\partial t} \left[
   \begin{array}{cc}
   u_k \\
   \\
   v_k \end{array} \right] =- \left[
   \begin{array}{cc}
   0 & L_I^{-1} \\
   \\
   1 & 0 \end{array} \right] \left[
   \begin{array}{cc}
   u_h\left( x_k + \frac {h_j}2 \right) \\ 
   \\
   v_h\left( x_k + \frac {h_j}2 \right) \end{array} \right]
   & &\\ 
  -\left[
  \begin{array}{cc}
   0 & 1 \\
   \\
    1 & 0 \end{array} \right]
   \left[
   \begin{array}{cc}
   u_h\left( x_k - \frac {h_j}2 \right) \\
   \\
   v_h\left( x_k - \frac {h_j}2 \right)
   \end{array} \right]
   -\left[
   \begin{array}{cc}
   \frac \gamma 2 \left(h_i + h_j\right) + \frac {h_j}2(\alpha u_k+\sin A_k)\\
   \\
   0
   \end{array} \right] \; .
   \end{eqnarray}
   which in terms of $A$ yields
   \begin{eqnarray}
   \label{shij}
   \left \{
   \begin{array}{cc}
   \displaystyle{\frac {{\rm d} A_k}{{\rm d} t}= u_k} & \\
   \\
   \displaystyle{\frac {{\rm d} u_k}{{\rm d} t}=
   \frac 1{C_I\frac{h_i}2+\frac{h_j}2}}&
   \displaystyle{
   \left(\frac{A_{k+1}-A_k}{h_j}-\frac{A_k-A_{k-1}}
   {L_Ih_i}-\frac{h_j}2\left( \alpha u_k+\sin A_k\right) -\gamma
   \left( \frac{h_i+h_j}2\right) \right)\; .}
   \end{array}
   \right.
   \end{eqnarray}
    
   {\bf APPENDIX B: The finite volume formulation in 2D}
   
   Here we follow a similar procedure as in the 1D case. 
   The  discretization of the domain in the x direction is the same 
   as done in the appendix A and in y direction we introduce the 
   fellowing partition
   \[
   W^{i,j}_{k}=\left \{ y \,| \,y_k-\frac{h_{i,j}'}2 \leq y < y_k +  \frac{h_{i,j}'}2\right \}
   \, \; ,
   \]
    where $y_k$ are the collocation points defined by
   \begin{eqnarray}
   \nonumber
   y_k&=&(k-\frac 12)h_i' \,\, {\rm in ~~passive~~ region}\; ,\\ \label{yk}
   y_k&=&(k-1)h_j'  \,\, {\rm in~~junction~~ region} \; .  \nonumber
   \end{eqnarray}
   Here $h_i'$ and $h_j'$ are the space mesh-length along y direction, 
   defined by 
   \[ h_i'=\frac{w'}{n_i'+\frac 12} \; ,h_j'=\frac{w}{n_j'-1} \; ,
   \]
   where $n_i'$ and $n_j'$ respectively are the number of discretization
   points in the passive and junction domains. The total number points 
   in the whole domain is $(2n_i + n_j) \times (2n_i' + n_j')$.
   
   For convenience we choose the same step size in the whole domain 
   (the number of points $n_i,n_j,n_i',n_j'$ are chosen such 
   $h_i=h_j=h_i'=h_j'=h$). Then the discretization of the domain can
   be written as
   \begin{displaymath}
   \Omega = \bigcup_{k,k'} \Omega_{k,k'} \, ,
   \end{displaymath}
   where
   \[
   \Omega_{k,k'} = \left \{ (x,y) \,| \,x_k-\frac{h}2 < x < x_k +  \frac{h}2
   \, \,{\rm and} \,\,  y_{k'}- \frac{h}2 < y < y_{k'} + \frac{h}2 \right \}  
   \, \; , \]
   are squares of side lenght $h$ centered at the point $(x_k,y_{k'})$.

   We integrate equation (\ref{wjj}) on each cell $\Omega_{k,k'}$
   and use the finite volume approximation to obtain
   \begin{equation}
   \label{int1}
   \dot \psi_{k,k'} \int_{\Omega_{k,k'}} C \,{\rm d}x {\rm d}y -
   \int_{\partial \Omega_{k,k'}} \frac 1 L \nabla \phi \cdot {\bf n} \, {\rm d} 
   \sigma  + (\alpha \psi_{k,k'} + \sin (\phi_{k,k'})) \int_{\Omega_{k,k'}} 
    \epsilon \,\,{\rm d}x {\rm d}y = 0 \; ,
   \end{equation}
   where $\psi = \dot \phi$, ${\bf n}$ is the unit exterior normal to 
   $\partial \Omega_{k,k'}$
   and $ \phi_{k,k'} = \phi(x,y)$, $ \psi_{k,k'} = \phi_t(x,y)$ are 
   assumed constant 
   for all $(x,y) \in \Omega_{k,k'}$. We then obtain {\bf for a cell inside
   the junction}
   \begin{equation}
   \label{sys}
   \left\{
   \begin{array}{cccc}
   \dot \phi_{k,k'} &=& \psi_{k,k'}\\ 
   \dot \psi_{k,k'} &=& \frac{1}{h^2}[\phi_{k+1,k'} +\phi_{k,k'+1} 
   +\phi_{k-1,k'} 
   + \phi_{k,k'-1} -4\phi_{k,k'}] +\sin \phi_{k,k'} + \alpha \psi_{k,k'} .
   \end{array} \right .
   \end{equation}
   In the system defined above $M$, $M'$ respectively are the total number 
   of points in the $x$ and $y$ directions. For $k=1,M$ and $k'=1,M'$, the 
   solution $(\phi_{k,k'},\psi_{k,k'})$
   is determined using the boundary condition (\ref{rbc}).     

   \noindent {\bf For a cell in the passive region} we obtain 
   \[
  \left \{
   \begin{array}{cccc}
   \dot \phi_{k,k'} &=& \psi_{k,k'}\\
   \dot \psi_{k,k'} &=& \frac{1}{h^2 L_I C_I}(\phi_{k+1,k'} +\phi_{k,k'+1}
   + \phi_{k,k'-1} -4\phi_{k,k'}) .
    \end{array}
    \right .
   \]

   Now consider cells that are overlapping the interface. We will
   dislpay two cases to show how the method works. 
   First consider {\bf the cell overlapping the bottom left corner 
   of the junction boundary} $(x_k= -l/2, y_k'=-w/2)$. 
   There (\ref{int1}) yields 
   \[
   \left \{
   \begin{array}{cccc}
   \dot \phi_{k,k'} &=& \psi_{k,k'}\\
   \dot \psi_{k,k'} &=& \frac{4}{h^2(3C_I +1)}[
   \phi_{k+1,k'}\frac{1}{2}(\frac{1}{L_I}+1) 
   +\phi_{k,k'+1}\frac{1}{2}(\frac{1}{L_I}+1)
   + \phi_{k-1,k'}\frac{1}{L_I} \\   & &
   + \phi_{k,k'-1}\frac{1}{L_I}
    -\phi_{k,k'}( 1 + \frac{3}{L_I}) ] 
   +\frac{1}{1+3C_I}(\sin \phi_{k,k'} + \alpha \psi_{k,k'}).
   \end{array}
   \right .
   \]
   Second consider {\bf a cell that is overlapping on the bottom 
   boundary of the junction }
   $(-l/2 < x_k < l/2, y_k'=-w/2)$. In that case we obtain 
   \[
   \left \{
   \begin{array}{cccc}
   \dot \phi_{k,k'} &=& \psi_{k,k'}\\
   \dot \psi_{k,k'} &=& \frac{2}{h^2(C_I +1)}[
   \phi_{k+1,k'}\frac{1}{2}(\frac{1}{L_I}+1)
   +\phi_{k,k'+1}
   + \phi_{k-1,k'} \\   & &
   + \phi_{k,k'-1}\frac{1}{L_I}
    -\phi_{k,k'}( 2 + \frac{2}{L_I}) ]
   +\frac{1}{1+C_I}(\sin \phi_{k,k'} + \alpha \psi_{k,k'}).
   \end{array}
   \right .
   \]
   
    With this formulation we have checked the conservation of energy
    in the absence of current or damping
    $\alpha =I=0$
    \[
    E = \int_{\Omega } [
   \frac {C}2 \left(\frac {\partial \psi}{\partial t}
   \right )^2 + \frac 1{2L} \vert\nabla\; \psi\vert^2 +
   (1-\cos(\phi)) \;] {\rm d}x{\rm d}y \; .
   \]
   For a number of points in $x$ and $y$, $M=100$, we
   obtain that the energy is conserved up to $10^{-3}$ in relative
   error for $t < 1 $.
   \newpage

   \begin{center} {\bf FIGURE CAPTIONS} \end{center}

   \vspace{.2in}

   \begin{enumerate}

   \item [1] Top panel: a view of a window Josephson junction. The
   bottom left panel shows a schematic top view. For the system shown
   on the right the linear region exists only on the left and right sides
   of the junction.
   
   \item [2] Phase plane $(X,u)$ for an antikink wave propagating across
   an interface at $x=0$. The solid line presents the solution of the
   adiabatic equations (\ref{odez}-\ref{odev}) and the losanges are the numerical 
   solution of the partial differential equation (\ref{a0f}). The 
   parameters are $\epsilon_0 = 0.6, \gamma = 0.02, L_I = 1$ 
   (top panel) and $L_I = 1.5$ (bottom panel).
 
   \item [3] Interface between a nonlinear and linear medium, the top 
   panel shows a plot of the antikink position $X(t)$ vs. time
   for the solution of the adiabatic equations (\ref{odez}-\ref{odev}) in full line.
   The crosses indicate the solution of the partial differential 
   equation (\ref{a0f}). The bottom panel presents the velocity $v(t)$.
   The parameters are $\epsilon_0 = 0.1 , \gamma = 0.01, L_I = 1$.

   \item [4] Simplified description of the fluxon motion in a one
   dimensional window junction for $L_I \gg 1$ (top panel) and
   $L_I \ll 1$ (bottom panel). The stable fixed points are shown
   by the large circles and the unstable ones are the small circles.

   \item [5] Motion of a fluxon in a one dimensional window junction
   from the numerical solution of (\ref{a0f}). The top and middle
   panels show respectively the position $X(t)$ and velocity $u(t)$
   vs. time while the bottom panel shows the phase plane $(X,u)$. The
   parameters are $\epsilon_0= 0, w'=2, L_I = C_I = 2, I = 0.1. $.
   
   \item [6] Contour plot of the phase $\varphi(x,y,t)$ for 
   $t=0, 400, 600 ,800$ of the numerical solution 
   of the two dimensional window junction equations
   (\ref{wjj}). The parameters are the same as in Figure 5.

   \item [7]  Contour plots of the phase $\varphi(x,y,t)$, numerical 
   solution of (\ref{wjj}) for $t=0,5, 6$ and 8 clockwise starting 
   from the top left panel. The parameters are $w'=7, L_I = C_I =1, 
   I=0.3$.
   
   \item [8]  Plots of the phase $\varphi(x,y,t)$, numerical
   solution of (\ref{wjj}) for $t=2, 10$ and 20 from left 
   to right. The top panels show the contours while the
   bottom show the three dimensional plots. The parameters are
   $w'=2, L_I = 0.0001, C_I =1, I=0.3$.

   \item [9]  Contour plots of the phase $\varphi(x,y,t)$, numerical
   solution of (\ref{wjj}) for $t=50100$ (top) and 51000 (bottom) 
   from left to right.
   The parameters are the same as in Fig. 8 except $L_I=10000$.

   \item [10]  Parameter plane $(L_I, C_I)$ showing the regions
   of existence of zero field steps (ZFS) corresponding to 
   the shuttling motion of a fluxon. The velocities $v_I$ are indicated
   in the top right corner of the picture.

   \end{enumerate}

   
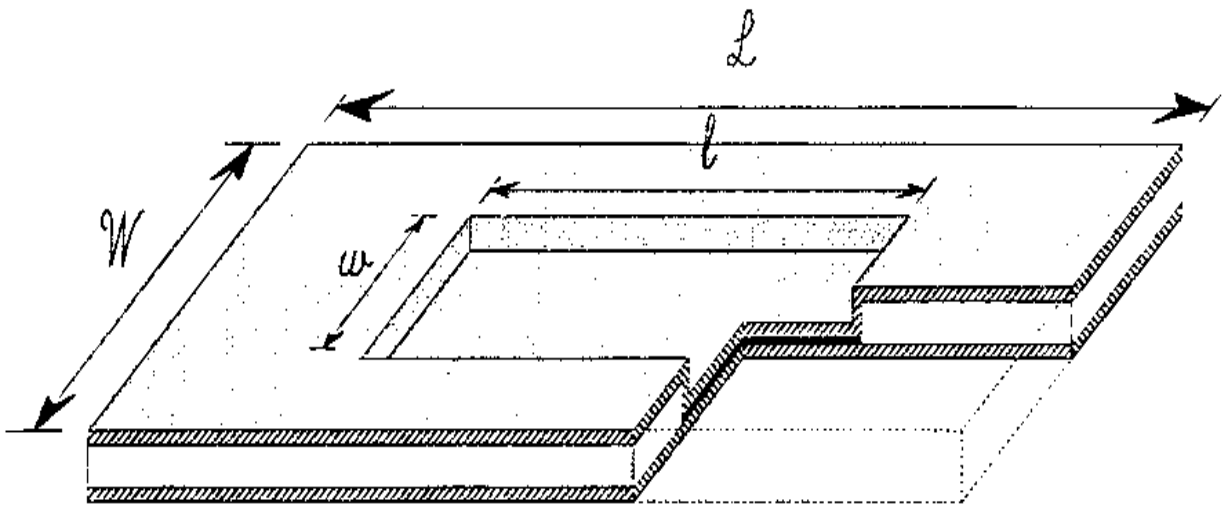
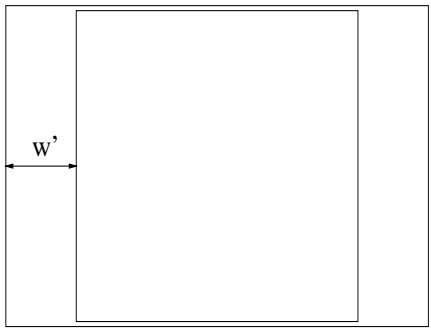
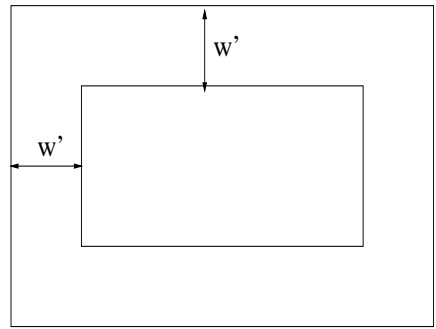
\end{document}